\documentclass[useAMS,twocolumn]{mn2e}
\usepackage{amsmath}
\usepackage{graphicx}
\usepackage{bm}

\def\vect{\boldsymbol}

\def\simlt{\lower.5ex\hbox{$\; \buildrel < \over \sim \;$}}
\def\simgt{\lower.5ex\hbox{$\; \buildrel > \over \sim \;$}}

\def\mag{\mbox{mag}}

\def\km{\mbox{km}}

\def\MHz{\mbox{MHz}}
\def\Gpc{\mbox{Gpc}}

\def\kpc{\mbox{kpc}}

\def\msun{\mbox{M}_\odot}

\def\deg{^\circ}

\def\arcmint{\mbox{arcmin}}

\def\21cm{\mbox{21-cm}} 

\def\mnras{MNRAS}
\def\apj{ApJ}

\def\apjs{ApJS}

\def\aap{AAP}

\def\nat{Nature}

\title{Imaging the Cosmic Matter Distribution using Gravitational Lensing of Pregalactic HI}

\date{\today}

\author[Hilbert, Metcalf \& White]{Stefan Hilbert\thanks{\texttt{hilbert@mpa-garching.mpg.de}}, R. Benton Metcalf  
and S. D. M. White\\ Max Plank Institut f\"ur Astrophysics,  
Karl-Schwarzchild-Str. 1, 85741 Garching, Germany}

\begin{document}

\maketitle

\begin{abstract}
\21cm emission from neutral hydrogen during and before the epoch of cosmic reionisation is gravitationally lensed by material at all lower redshifts. Low-frequency radio observations of this emission can be used to reconstruct the projected mass distribution of foreground material, both light and dark. We compare the potential imaging capabilities of such \21cm lensing with those of future galaxy lensing surveys. We use the Millennium  Simulation to simulate large-area maps of the lensing convergence with the noise, resolution and redshift-weighting achievable with a variety of idealised observation programmes. We find that the signal-to-noise of \21cm lens maps can far exceed that of any map made using galaxy lensing. If the irreducible noise limit can be reached with a sufficiently large radio telescope, the projected convergence map provides a high-fidelity image of the true  matter distribution, allowing the dark matter halos of individual galaxies to be viewed directly, and giving a wealth of statistical and morphological information about the relative distributions of mass and light. For instrumental designs like that planned for the Square Kilometer Array (SKA), high-fidelity mass imaging may be possible near the resolution limit of the core array of the telescope.
\end{abstract}


\section{introduction}

Since~\nocite{Zwicky33}{Zwicky} (1933) first realised that unseen material is needed to
explain the dynamics of galaxy clusters, many observations have
indicated that large-scale structures are dominated by some form of
dark matter.  The now widely accepted cold dark matter (CDM) model
provides a consistent explanation for cosmic microwave background
(CMB) fluctuations, for type Ia supernova distances, for clustering
measures from galaxy redshift surveys, for galaxy cluster abundances
and their evolution, and for the statistics both of weak gravitational
lensing and of Ly$\alpha$ absorption in quasar spectra. The universe
apparently contains about five times as much dark matter as ordinary
baryons, providing in total about a quarter of the closure density.
According to the CDM picture, every galaxy has its own dark matter
halo, which may be partially disrupted in a group or cluster to
produce a common halo.  These structures can be predicted in great
detail by numerical simulations, but the predictions are yet to be
convincingly verified because we are unable to map the dark matter
distribution in enough detail to make a proper comparison.  Weak
gravitational lensing of distant galaxies has allowed progress to be
made, but only near the centres of the largest galaxy clusters is the
signal-to-noise sufficient for true mapping. The limitations of this
approach are clearly indicated by the recent image of a representative
field made using deep Hubble Space Telescope data by
\nocite{Massey07}Massey {et~al.} (2007b). The resolution and sensitivity of lensing maps based
on galaxies are fundamentally limited by the finite number density and
the intrinsic ellipticities of the sources.  In this paper, we
demonstrate that much higher fidelity and resolution can be achieved
if future observations allow pregalactic HI to be used as the
gravitationally lensed source.

The spin temperature of neutral hydrogen during and before the epoch of
reionisation ($8 \simlt z \simlt 300$) fell out of thermal equilibrium  
with the CMB radiation, resulting in the absorption and emission of \21cm
radiation.  There has been a great deal of interest in the prospect of
detecting and mapping this radiation using radio telescopes now under
construction or in planning \nocite{astro-ph/0608032}(see Furlanetto, Oh \&  Briggs 2006, for an extensive
review).  This radiation provides an excellent source  
for gravitational lensing studies.  Structure is expected in the \21cm  
emission down to arcsecond scales, and at each point on the sky there will be  
$\sim 1000$ statistically independent regions at different redshifts, and thus
frequencies, that could in principle be observed.  Gravitational  
lensing will coherently distort the \21cm brightness temperature maps at these  
different frequencies. For each frequency, the gradient in the brightness temperature may be used to obtain an estimate of the lensing distortion. Since the intrinsic structure in the HI gas that acts as noise on the estimate is uncorrelated for maps at (sufficiently) different frequencies, the coherent distortion of the brightness temperature can be measured with high accuracy if enough independent redshifts are observed. In this way, a map of the
foreground matter density can be constructed \nocite{ZandZ2006,metcalf&white2007}({Zahn} \& {Zaldarriaga} 2006; Metcalf \& White 2007).

Observing the \21cm radiation at high redshift is challenging. Foregrounds of atmospheric, galactic, or extragalactic origin (e.g. synchrotron radiation from electrons) dominate over the \21cm signal in the relevant frequency range \nocite{astro-ph/0608032}(Furlanetto, Oh \&  Briggs 2006). The foregrounds are expected to vary much less with frequency (and with position on the sky for galactic foregrounds) than the \21cm signal from high-redshift HI structures \nocite{ZFH2004,ZandZ2006}({Zaldarriaga}, {Furlanetto}, \& {Hernquist} 2004; {Zahn} \& {Zaldarriaga} 2006). Therefore, it is hoped that the foreground radiation and  the \21cm signal may be separated by modelling the foregrounds as slowly varying functions of frequency.

Subtracting foregrounds from the observed radiation will be complicated 
and will contribute noise to the temperature map.
In addition to the noise from foreground residuals, there is {\it
irreducible noise} in the mass map constructed from the  \21cm-lensing signal, 
which comes from the unknown intrinsic structure of the \21cm brightness  
temperature distribution.  This noise cannot be reduced by increasing the  
collecting area of the telescope, by increasing the integration time or by improving the
removal of foregrounds.  \nocite{metcalf&white2007}Metcalf \& White (2007) showed that if the
signal-to-noise in the brightness temperature map at each frequency is  
greater than one, then the noise in the mass map will be close to the  
irreducible value.  Increasing the frequency resolution of the radio observations
increases the number of effectively independent regions along the
line-of-sight until the bandwidth becomes smaller than the radial
correlation length of structure in the brightness temperature  
distribution.
If the bandwidth is matched to the correlation length, the irreducible  
noise is minimised.  The correlation length in turn depends on beam size, and is
smaller for smaller beams.  Thus unlike galaxy lensing surveys, the
irreducible noise {\it decreases} with increasing resolution for \21cm
lensing.  In practise, there is a trade-off because smaller bandwidth  
means less flux, but this can be compensated by increasing collecting area  
and/or integration time.  In this paper, we study what is achievable with an
idealised radio telescope, so we assume that the irreducible noise  
level is reached, using values calculated by \nocite{metcalf&white2007}Metcalf \& White (2007) as a  
function of beam-size and frequency.

At least in principle, a \21cm lensing survey will be much less noisy  than surveys using galaxies (because of the larger effective number of  sources) and will have a substantially stronger signal (because of the greater distance of the sources and the additional structure that lies in front of them). 
For a galaxy shear map, the noise increases with decreasing smoothing because fewer galaxies are used to estimate the shear at each point of the map.  The opposite is true for \21cm lensing, where a smaller beam allows one to observe more independent sources along each line-of-sight. It should therefore be possible to  make high-resolution images of the matter distribution at high  signal-to-noise using \21cm lensing, while the smallest scale over which galaxy lensing  can map with $S/N>1$ is $\simgt 1\,\arcmint$, even using an ambitious dedicated space telescope. In addition,  \21cm lensing will provide information about the mass distribution at redshifts much higher than can be probed by galaxy lensing. The results we show below illustrate these points clearly.

This paper is organised as follows.  In Sec.~\ref{sec:lensing}, the  relevant elements of lensing theory are introduced, and the parameters of our idealised surveys are discussed. Our lensing simulation method is described in Sec.~\ref{sec:simulations}. The results of  our simulations are presented in Sec.~\ref{sec:results}. Sec.~\ref{sec:conclusion} contains our conclusions.

\section{lensing preliminaries}\label{sec:lensing}

Gravitational lensing shifts the observed position of each point in the image of a distant source.  Take the observed angular position on the sky to  be $\vect{\theta}$ and the position in the absence of lensing to be $\vect{\beta}$. The first-order distortion in the image is expressed by the derivatives of the mapping between these angles. The distortion matrix is commonly decomposed into the convergence $\kappa$ and two components of shear, $\bf \gamma$, defined by
\begin{equation}
\left[ \frac{\partial \vect{\beta}}{\partial \vect{\theta}}\right] = \left(
\begin{array}{ccc}
1-\kappa +\gamma_1  & \gamma_2   \\
\gamma_2  &  1-\kappa-\gamma_1
\end{array}
\right).
\end{equation}

To lowest order and to an excellent approximation \nocite{2003ApJ...592..699V}({Vale} \& {White} 2003), the convergence is related directly to the distribution of matter  
through
\begin{subequations}
\label{eq:convergence}
\begin{align}
\kappa\left(\vect{\theta}\right)
&=
\frac{3}{4} H_o \Omega_m \int_0^\infty dz ~ \frac{(1+z)^2}{E(z)} g\left(z \right) \delta\left(D(z)\,\vect{\theta},z\right)
\\  & \simeq 
\frac{4\pi G}{c^2} \sum_i  g\left(z_i\right) \left(\Sigma_i\left(\vect{\theta}\right)  - \overline{\rho}\left(z_i\right) \delta l_i \right)  =  \sum_i \kappa_i \label{multiplane}
\end{align}
with
\begin{equation}
g(z) =  (1+z)^{-1} \int_{z}^\infty dz' ~ w\left(z'\right)  
\frac{D(z,0)D(z',z)}{D(z',0)}.
\end{equation}
\end{subequations}
Here $D(z',z)$ is the angular size distance between the two redshifts, and $\delta(\vect{x},z)$ is the fractional density fluctuation at redshift $z$ and perpendicular position $\vect{x}$.  The function  $E(z)=\sqrt{\Omega_m(1+z)^3 + \Omega_\Lambda + (1-\Omega_m-\Omega_\Lambda)(1+z)^2}$, where $\Omega_m$ and $\Omega_\Lambda$, are the densities of matter and the cosmological constant measured in units of the critical density. The weighting function for the source distance distribution, $w(z)$, is normalised to unity.

Equation~\ref{multiplane} is the multiple-lens-plane approximation, in which $z_i$ is the redshift of the $i$th lens plane, $\Sigma_i\left(\vect{\theta}\right)$ is its surface density, $\delta l_i$ is its proper thickness, and $\overline{\rho}(z_i)$ is the average matter  density of the universe.  This approximation is well justified if the planes are thin compared to the range in redshift over which $g(z)$ varies and $\kappa_i\simgt 1$ for no more than one lens plane.  This second  requirement is well justified for all but a very small fraction of the sky where multiple galaxy clusters happen to overlap in projection.

When we consider galaxy surveys, we model the redshift distribution of usable galaxies as
\begin{equation}
\label{eq:gal_z_distr}
 w(z)=\frac{3z^2}{2z_0^3}\exp\left[-\left(\frac{z}{z_0}\right)^{3/2} \right]\,\mbox{, where }\,
z_0=\frac{z_\mathrm{med}}{1.412}
\end{equation}
is set to obtain the median redshift $z_\mathrm{med}$ appropriate for each specific survey \nocite{1994MNRAS.270..245S}({Smail}, {Ellis} \&  {Fitchett} 1994).  We estimate the smoothed convergence distribution for a Gaussian smoothing kernel defined by
\begin{equation}
\label{eq:df_Gauss_kernel}
W(\theta) = \frac{2}{\pi\lambda^2} \exp\left(-\frac{2\theta^2}{\lambda^2}  
\right),
\end{equation}
where $\theta$ denotes the angular separation between two points on the sky, and the `beam diameter' $\lambda$ quantifies the spatial scale of the smoothing.
For the kernel~\eqref{eq:df_Gauss_kernel}, the correlation function for the noise in the convergence map is given by
\begin{equation}\label{galaxy_noise}
\xi_\mathrm{N}(\theta)  = \frac{\sigma^2_\epsilon}{2\pi\lambda^2n_\mathrm{g}}
\exp\left( -\frac{\theta^2}{\lambda^2}\right)\;,
\end{equation}
where $n_\mathrm{g}$ is the number density of source galaxies on the sky, and $\sigma_\epsilon$ is the standard deviation in the magnitude of their ellipticities \nocite{2000MNRAS.313..524V}({van Waerbeke} 2000). A realistic value is $\sigma_\epsilon = 0.3$ [for example, $\sigma_\epsilon = 0.32+0.0014(\mag-20)^3$ for the ACS camera on HST \nocite{Massey07b}(Massey {et~al.} 2007a)]. The proposed satellite SNAP\footnote{snap.lbl.gov} is expected to  survey $\sim 2\%$ of the sky with an estimated galaxy density of  $n_\mathrm{g}\simeq100\,\arcmint^{-2}$ and a median redshift $z_\mathrm{med}\sim 1.23$. [For  comparison, \nocite{Massey07}Massey {et~al.} (2007b) were able to use 71 galaxies per square arcminute in the HST COSMOS survey.] The DUNE\footnote{www.dune-mission.net} satellite proposes surveying $\sim  50\%$ of the sky with $n_\mathrm{g}\simeq 35\,\arcmint^{-2}$ and a median redshift $z_\mathrm{med}\sim 0.9$. Several proposed ground based surveys -- LSST\footnote{www.lsst.org},  PanSTARRS\footnote{pan-stars.ifa.hawaii.edu}, VISTA\footnote{www.vista.ac.uk} -- hope to reach source number densities comparable to DUNE. In the following, we adopt these two sets of  parameters as our optimistic assessment of the parameters defining future {\it  space-} and {\it ground-based} galaxy surveys. For a Gaussian smoothing with  $\lambda=1\,\arcmint$, they yield a normalization,  $\sigma_\mathrm{N}=\sqrt{\xi_\mathrm{N}(\theta=0)}$, of 0.012 and 0.02,  respectively, for the noise correlation. A lower noise level 

When simulating convergence maps derived from \21cm observations, we will make the approximation $w(z)=\delta(z-z_0)$. This is reasonable because angular size distances are a weak function of source redshift over the relevant range.  The noise in the convergence map is worked out in \nocite{metcalf&white2007}Metcalf \& White (2007) under the assumption that the high-frequency components of maps of pregalactic HI decorrelate with increasing redshift separation in the same way as those of maps of the underlying cold dark matter distribution. In this case, the noise is very well approximated as a Poisson process smoothed by the telescope beam, which we again model as Gaussian.  This results in the same correlations as in Eq.~(\ref{galaxy_noise}) except with a different normalisation $\sigma_\mathrm{N}$. Here, we adopt normalisations of 0.0042 for a $\lambda=6$ arcsecond beam and 0.014 for a $\lambda=1$ arcminute beam. These values are representative for surveys that observe the \21cm radiation at redshifts around $z_0=12$, work close to the irreducible-noise limit, cover $\sim 10\,\MHz$ in frequency, and have optimal bandwidth $\sim 0.05\,\MHz$.
A beam size $\lambda=6\arcsec$ is very futuristic, since it corresponds to a densely filled array with baselines of order $100\,\km$. A beam with $\lambda=1\,\arcmint$ might be realized with the planned Square Kilometer Array\footnote{www.skatelescope.org} (SKA) \nocite{metcalf&white2007}(Metcalf \& White 2007). The assumed source redshift $z_0=12$ for the \21cm radiation lies within the expected obvservable redshift range of the SKA.

\section{simulations}\label{sec:simulations}

We simulated maps of the lensing convergence using the Millennium Simulation~\nocite{SpringelEtAl05MSReview}({Springel} {et~al.} 2005) and the multiple-lens-plane approximation. The Millennium Simulation is a very large N-body simulation of cosmological structure formation containing $10^{10}$ particles in a cubic region of  $L=500 h^{-1}\,\mbox{Mpc}$ comoving on a side.  The cosmological parameters for the simulation are: $\Omega_m=0.25$, $\Omega_\Lambda=0.75$, and a Hubble constant of $h=73$ in units of $100\,\mbox{km}\,\mbox{s}^{-1}\mbox{Mpc}^{-1}$. The initial density power spectrum is scale-invariant (spectral index $n=1$) with normalisation $\sigma_8=0.9$.  Snapshots of the matter distribution were stored on disk at 64 output times between redshift $z=127$ and $z=0$. For $0\le z \le 1$, the snapshots are spaced at roughly 200 Myr intervals resulting in 23 snapshots in that range. Above redshift unity, the snapshots are spaced approximately logarithmically in the expansion factor.

For each snapshot of the simulation, we project the matter distribution  in a slice of appropriate thickness onto a plane and place it at the  snapshot's redshift along the line of sight. The projected matter density on these  lens planes is represented by meshes with a spacing of $2.5 h^{-1}\,\kpc$ comoving. In order to reduce discreteness noise while retaining the high resolution of the simulation, an adaptive smoothing kernel is used.  Before projection, the mass associated with each particle is distributed in a spherical cloud with Gaussian density profile and {\it rms} radius equal to half the  distance to its 64th nearest neighbour. The projected density at each mesh point on the lens plane is then calculated by summing the  contributions from each particle.

To create convergence maps over a field of $10\deg\times10\deg$, we shoot $36000\times36000$ light rays through the series of 50 lens planes which extend from $z=0$ to $z=9$. On each lens plane, we calculate the projected matter density at the position of each ray by bilinear interpolation between surrounding mesh points. Using Eq.~\ref{eq:convergence}, the convergence $\kappa_i$ is then calculated and summed for each ray to get $\kappa(\vect{\theta})$ for the assumed source redshift distribution [i.e. $w(z)=\delta(z-12)$ for the \21cm emission, and $w(z)$ given by Eq.~\eqref{eq:gal_z_distr} for the galaxies].

The convergence maps obtained by this procedure have a resolution of 1 arcsecond and are essentially noise-free. In order to simulate maps as they would be observed, we add independent Gaussian noise with appropriate dispersion to each pixel and smooth the maps using a Gaussian filter representing either the radio telescope beam or the required smoothing in the case of galaxy lensing.  This procedure yields convergence maps with the desired resolution and with noise satisfying Eq.~(\ref{galaxy_noise}).

For a $10\deg\times10\deg$ field, the comoving volume of the backward light cone out to $z=12$ is more than 32 times that of the Millennium Simulation. The region out to $z=2$ which contains most of the lensing structures is still 4.5 times larger in comoving volume than the simulation.  Every simulated object thus contributes to the projected mass distribution several times.  As explained in \nocite{hilbert2007}{Hilbert} {et~al.} (2007), the lens planes were constructed by projection along a line-of-sight which is tilted with respect to the principal axes of the simulation, and as a result are periodic on a rectangular cell of size $1.58\times1.66h^{-2}\Gpc^2$ comoving; the periodic length normal to the lens planes is $5.24h^{-1}\Gpc$ comoving. There are objects that appear multiple times
at the same redshift when $z>1.2$, but the number of these cases becomes significant (i.e. exceeds 1/3 of all objects at a given redshift) in our field only for $z>2$.
Objects also appear multiple times at different redshifts. However, objects are projected on top of their own images only in very few (special) directions and for very widely separated redshifts.\footnote{
Our lens-plane geometry ensures that multiple copies at the same or similar redshift are separated by angles equal to or not much smaller than the angular scale of the simulation box at that redshift.
These multiple copies introduce artificial correlations on large angular scales and decrease the statistical independence of well-separated parts of the simulated field, but they do not affect correlations on smaller angular scales, nor do they alter one-point statistics such as the convergence probability distribution. Since the artificial correlations have different angular scales at different redshifts, multiple copies have different foregrounds and backgrounds in projection.
}
Generally multiple copies of objects are almost always seen at different redshifts and are almost always projected onto different foregrounds and backgrounds. As a result, there is effectively no duplication of projected structure within this field, despite the fact that the total comoving length of the line-of-sight out to $z=12$ is more than 14 times the side of the computational box

\section{results}\label{sec:results}

\subsection{Images}\label{sec:images}

\begin{figure*}
\centerline{
{\includegraphics[width=0.42\linewidth]{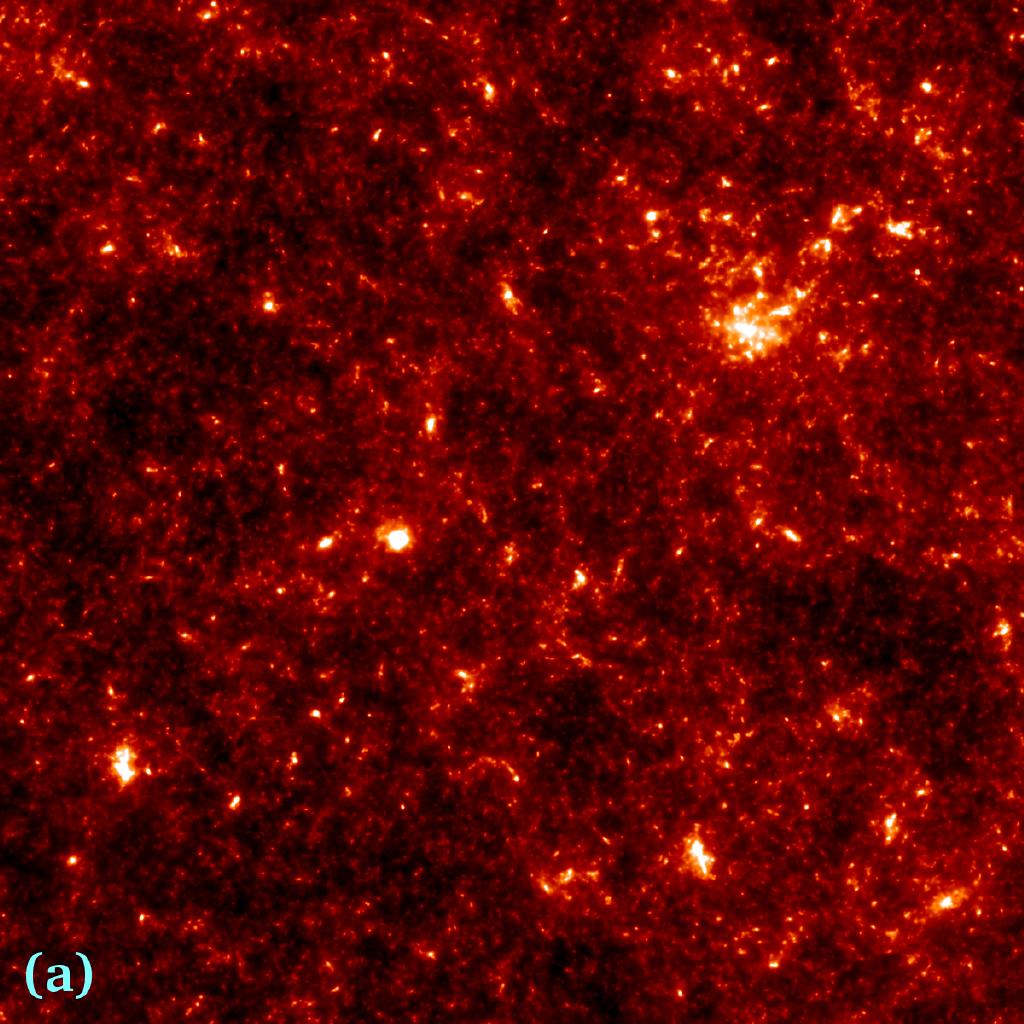}}
  \hspace{0.1em}
{\includegraphics[width=0.42\linewidth]{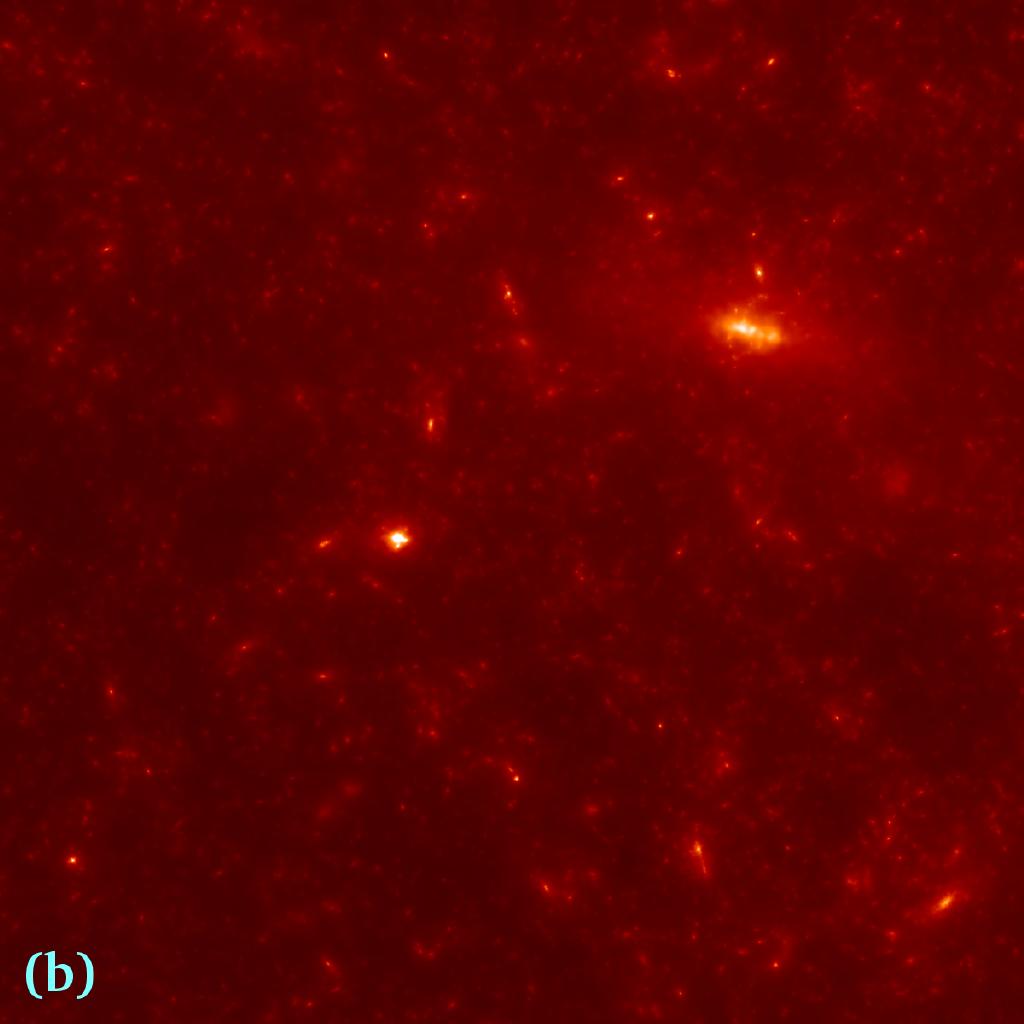}}
  \hspace{1em}
{\includegraphics[width=0.07\linewidth]{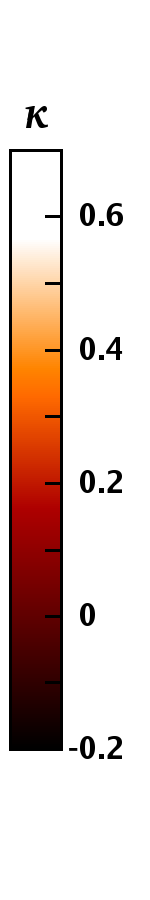}}
}
\vspace{0.2em}
\centerline{
{\includegraphics[width=0.42\linewidth]{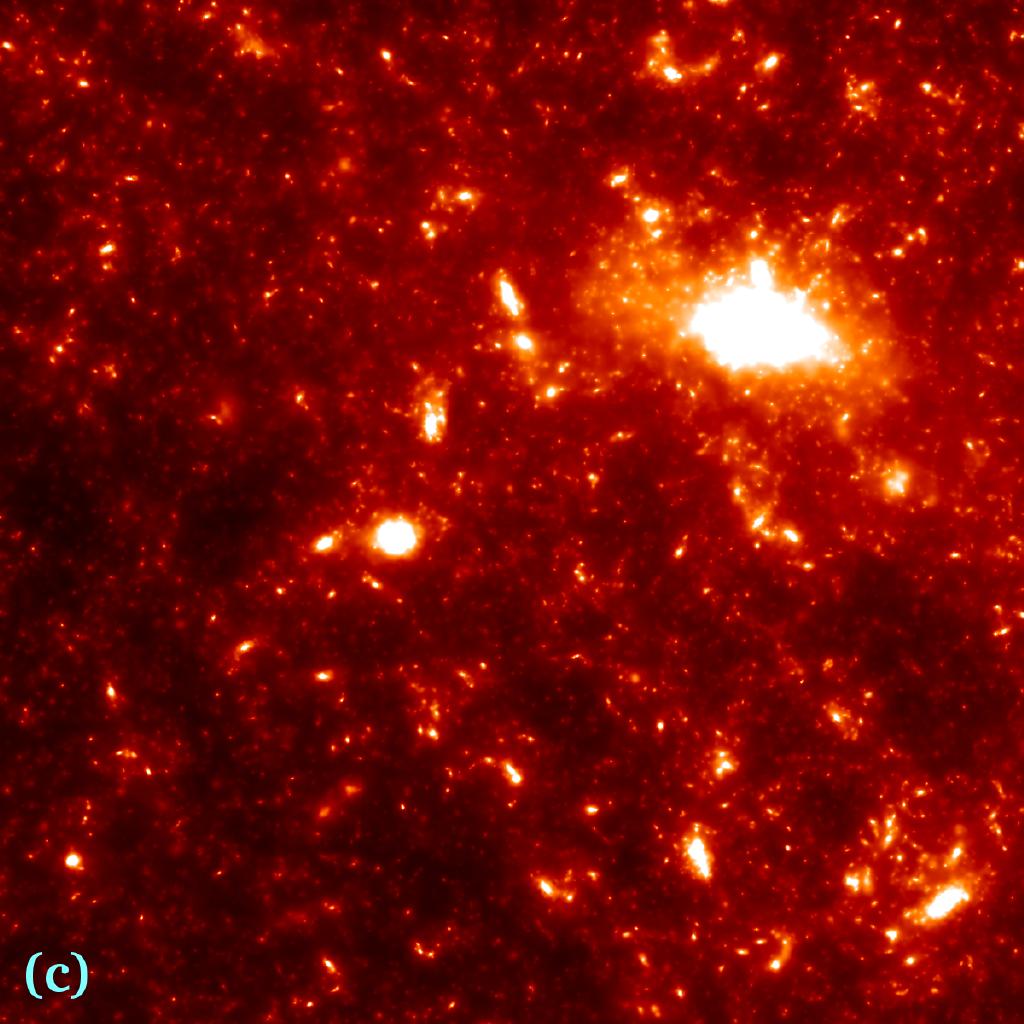}}
 \hspace{0.1em}
{\includegraphics[width=0.42\linewidth]{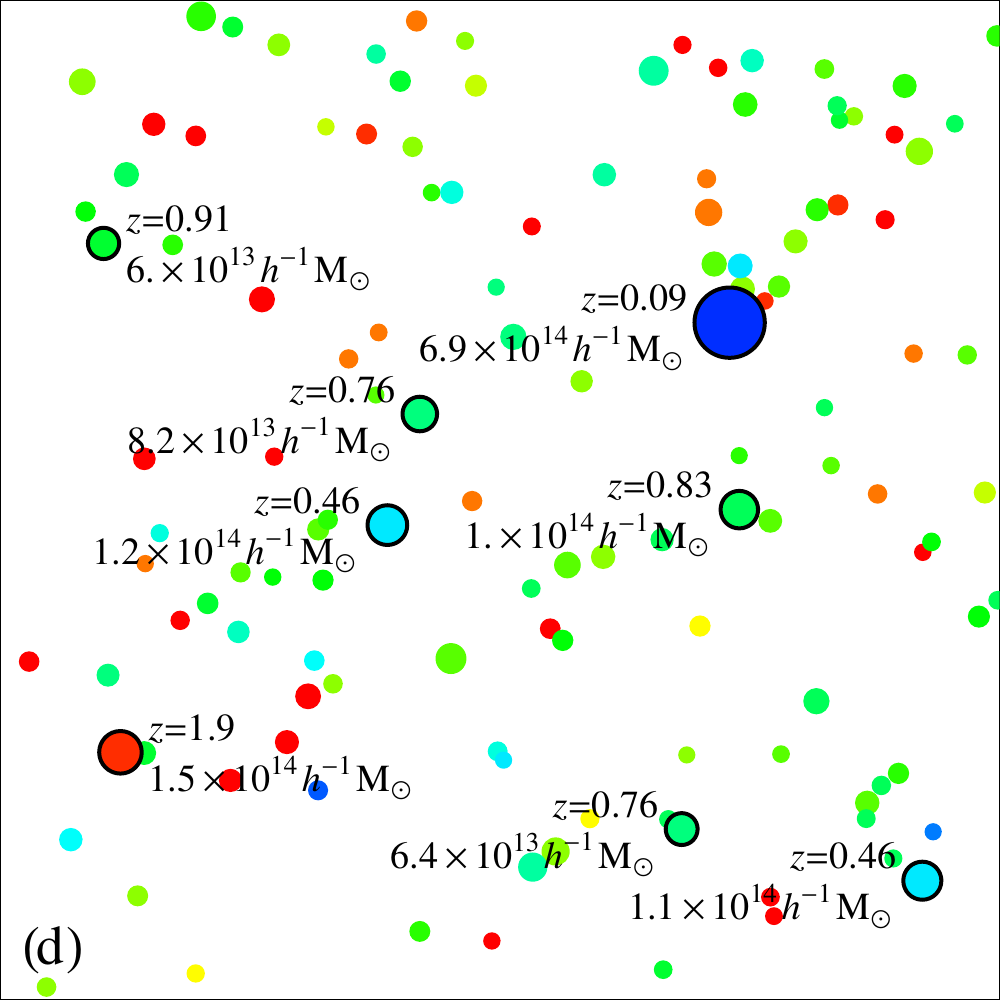}}
  \hspace{1em}
  \hspace{0.01\linewidth}
{\includegraphics[width=0.05\linewidth]{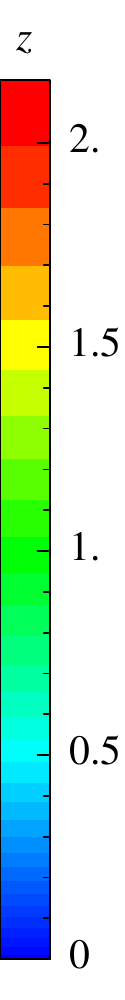}}
}
\caption{
  \label{fig:maps1} Convergence maps in a field of $20\arcmin\times20\arcmin$.
(a), (b) and (c) are all unsmoothed maps at $1\arcsec$ resolution. (a) has
the distance weighting appropriate for HI sources at $z=12$, while (b)
and (c) are weighted as appropriate for a space-based galaxy survey
with median redshift $z_\mathrm{med}=1.23$.  Both (a) and (b) follow the
colour scale indicated by the bar. (c) is identical to (b) but with
contrast enhanced by a factor of 11/3 to allow closer comparison with
(a). Panel~(d) identifies redshifts (colour-coded) and virial
masses (size-coded) for all objects in the field with
$M_{200}>10^{13}h^{-1}\,\msun$. The eight most massive objects
are labelled explicitly.}
\end{figure*}

\begin{figure*}
\vspace{0.2em}
\centerline{
{\includegraphics[width=0.42\linewidth]{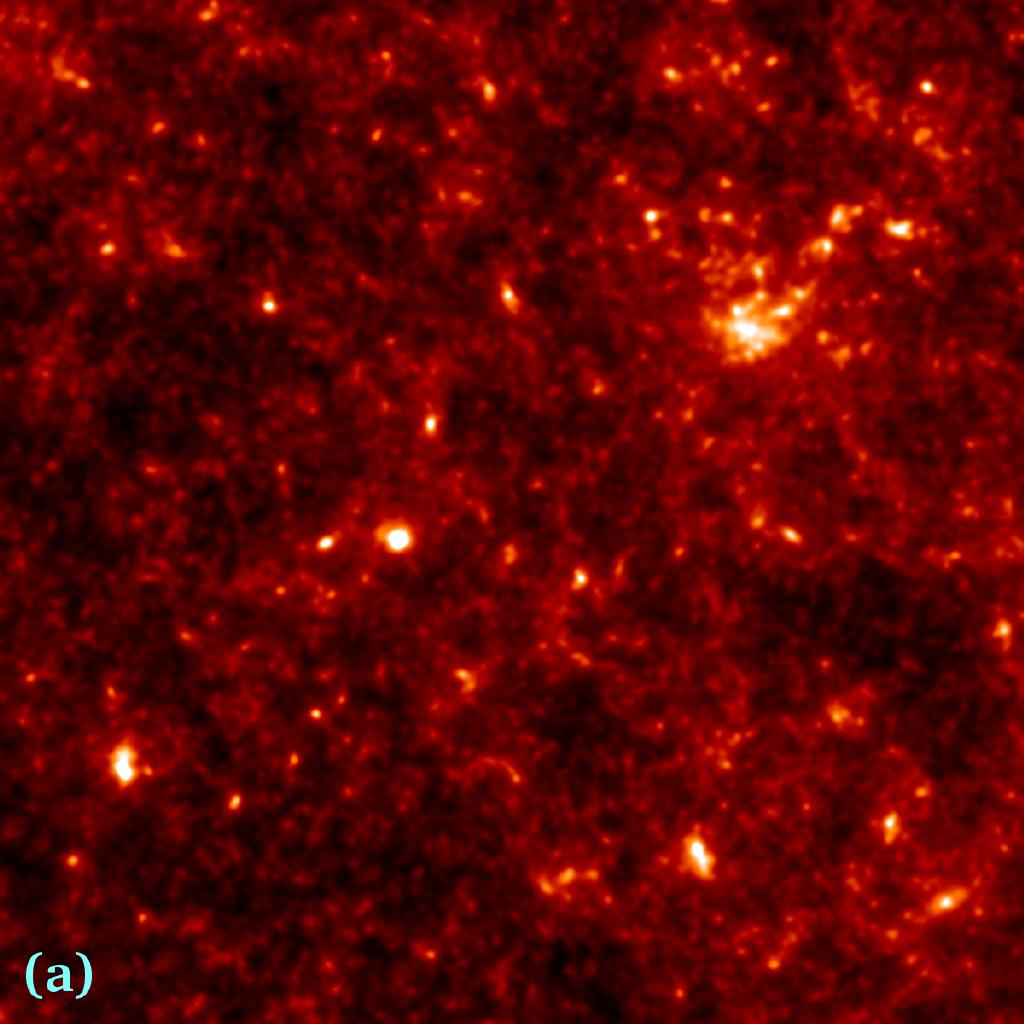}}
  \hspace{0.1em}
{\includegraphics[width=0.42\linewidth]{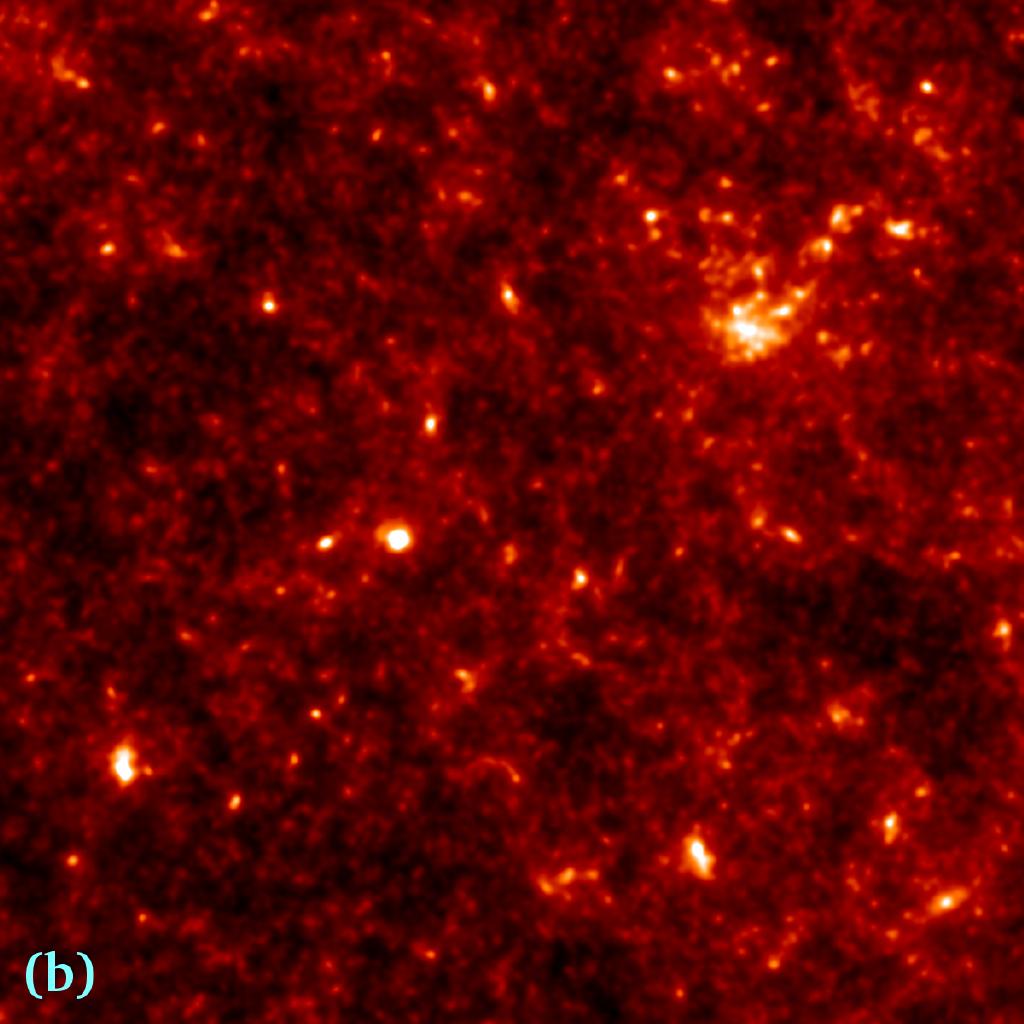}}
  \hspace{1em}
{\includegraphics[width=0.07\linewidth]{colorbar_kappa_1.png}}
}
\caption{
\label{fig:maps2}
\21cm-based convergence maps for the $20\arcmin\times20\arcmin$ field shown in Fig.~\ref{fig:maps1}(a), but smoothed assuming a telescope beam with $\lambda=6\arcsec$.
Whereas (a) is noise-free, noise has been added in (b) at the irreducible value for a map of this
resolution. The colour scale indicated by the bar at right is the same as in Fig.~\ref{fig:maps1}(a).
}
\end{figure*}

\begin{figure*}
\centerline{
{\includegraphics[width=0.42\linewidth]{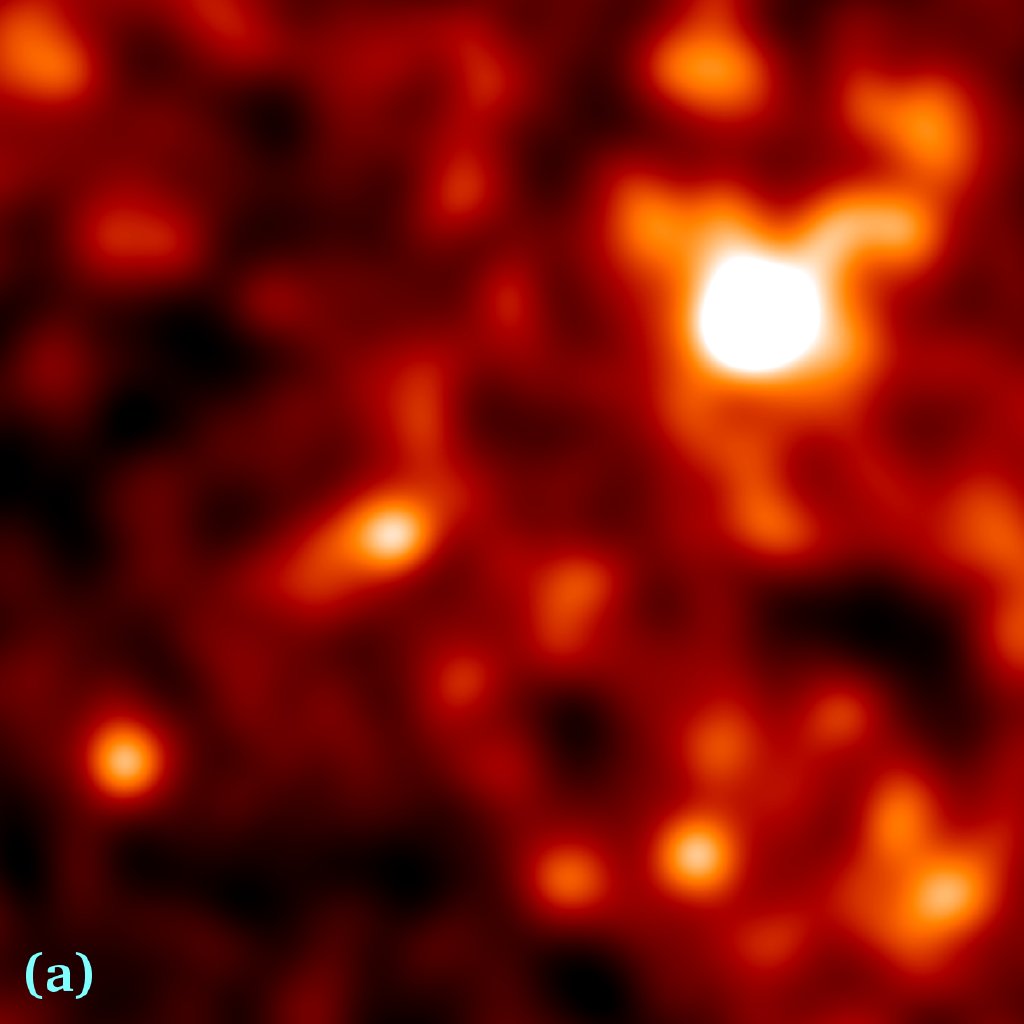}}
   \hspace{0.1em}
{\includegraphics[width=0.42\linewidth]{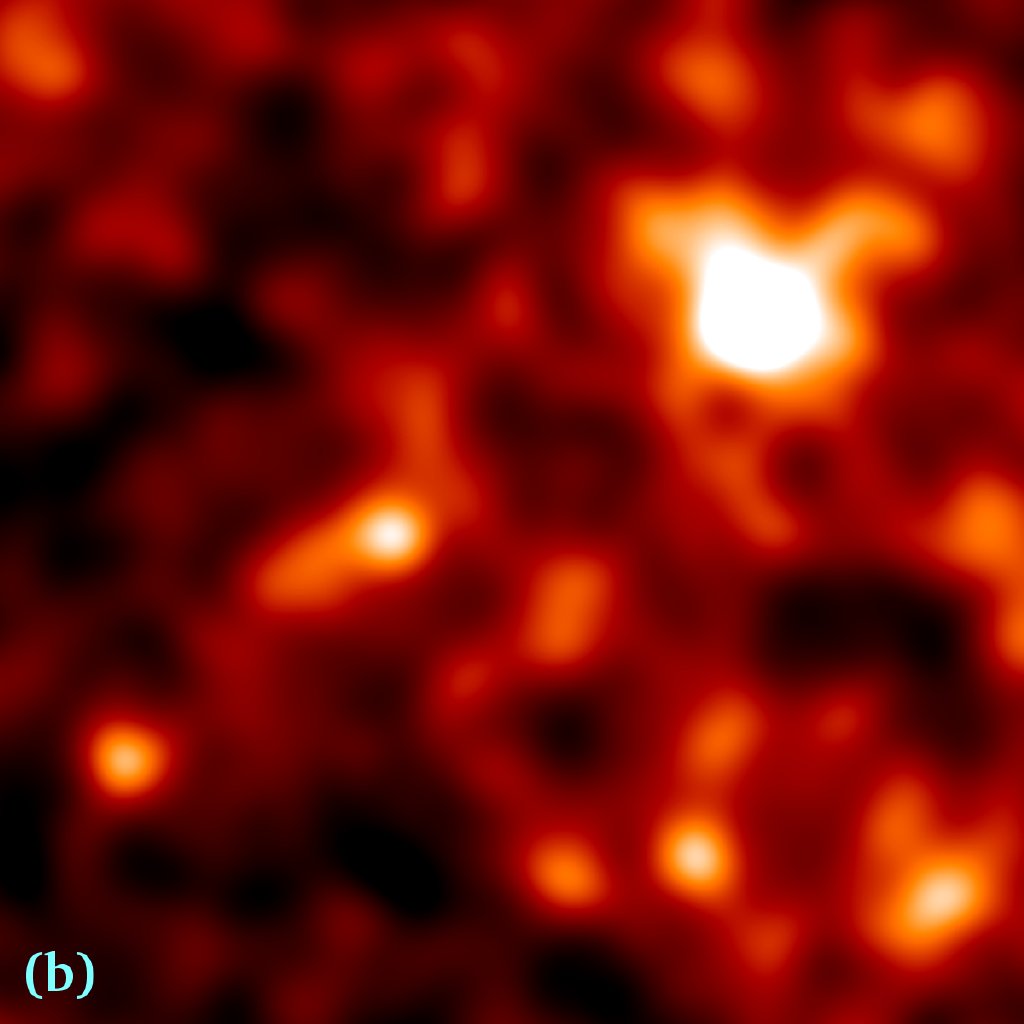}}
   \hspace{1em}
{\includegraphics[width=0.07\linewidth]{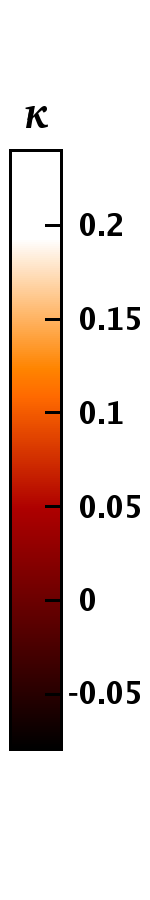}}
}
\vspace{0.1em}
\centerline{
{\includegraphics[width=0.42\linewidth]{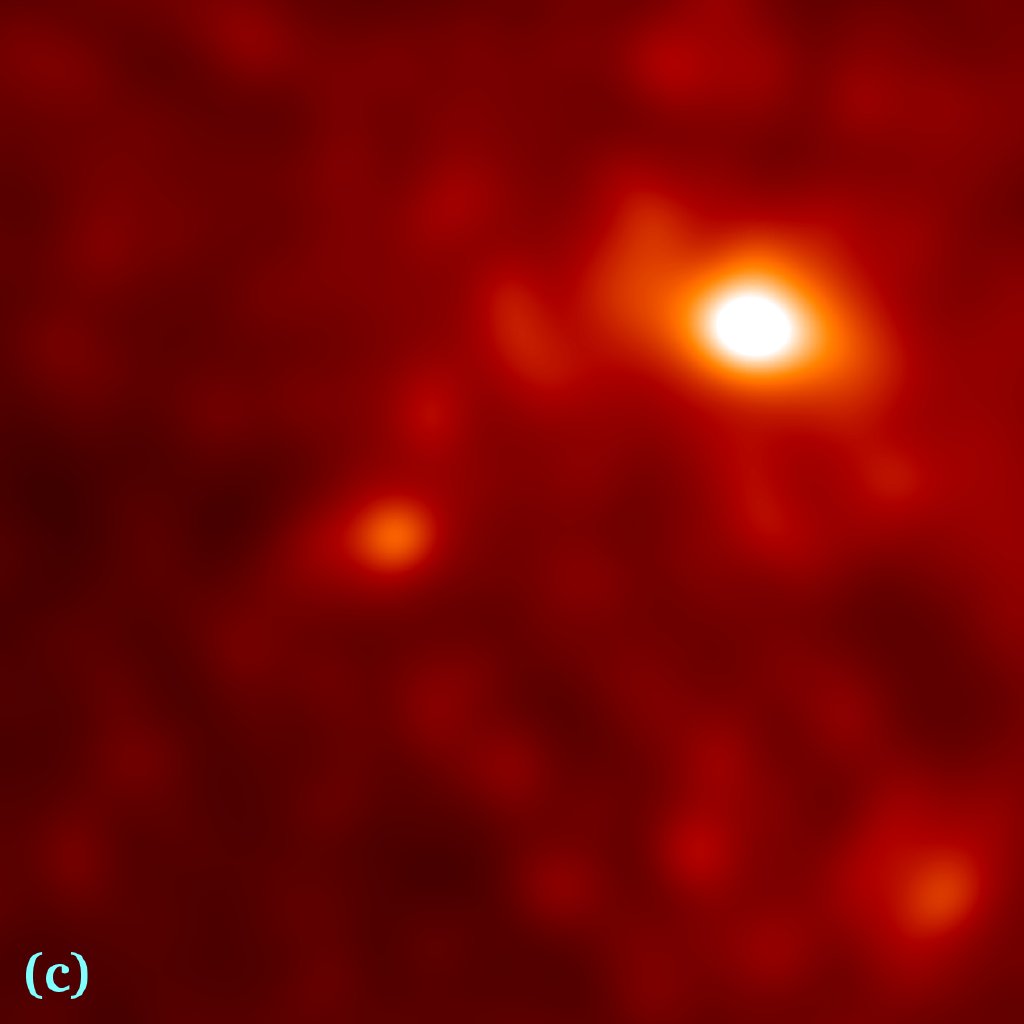}}
   \hspace{0.1em}
{\includegraphics[width=0.42\linewidth]{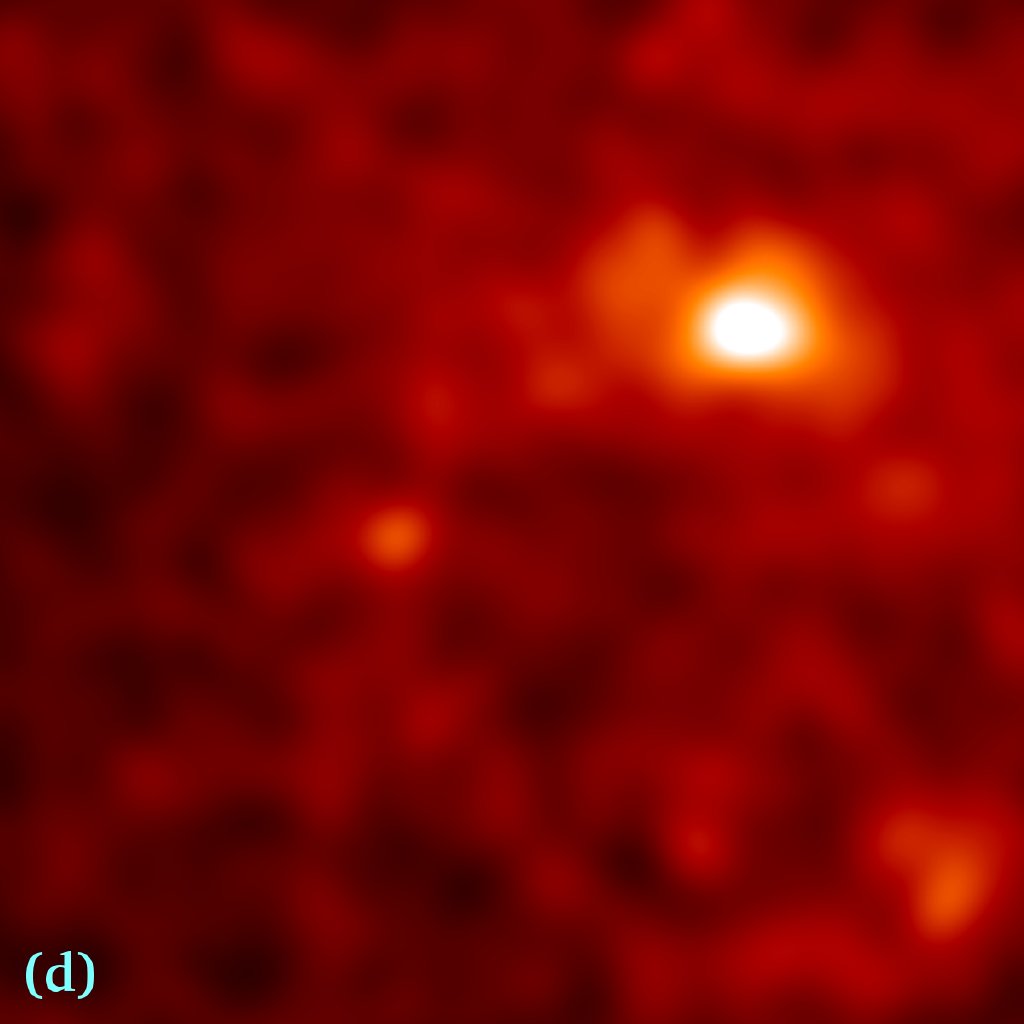}}
   \hspace{1em}
   \hspace{0.07\linewidth}
}
\vspace{0.1em}
\centerline{
{\includegraphics[width=0.42\linewidth]{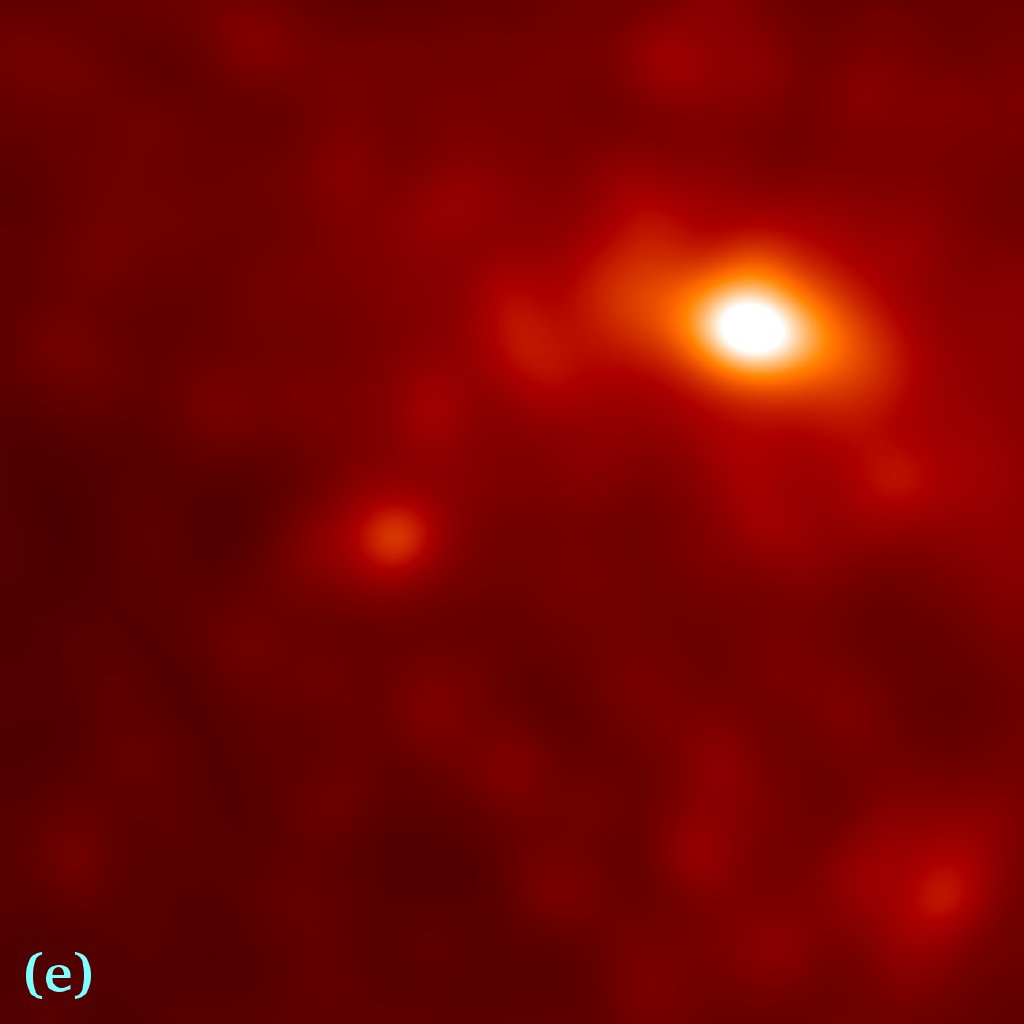}}
   \hspace{0.1em}
{\includegraphics[width=0.42\linewidth]{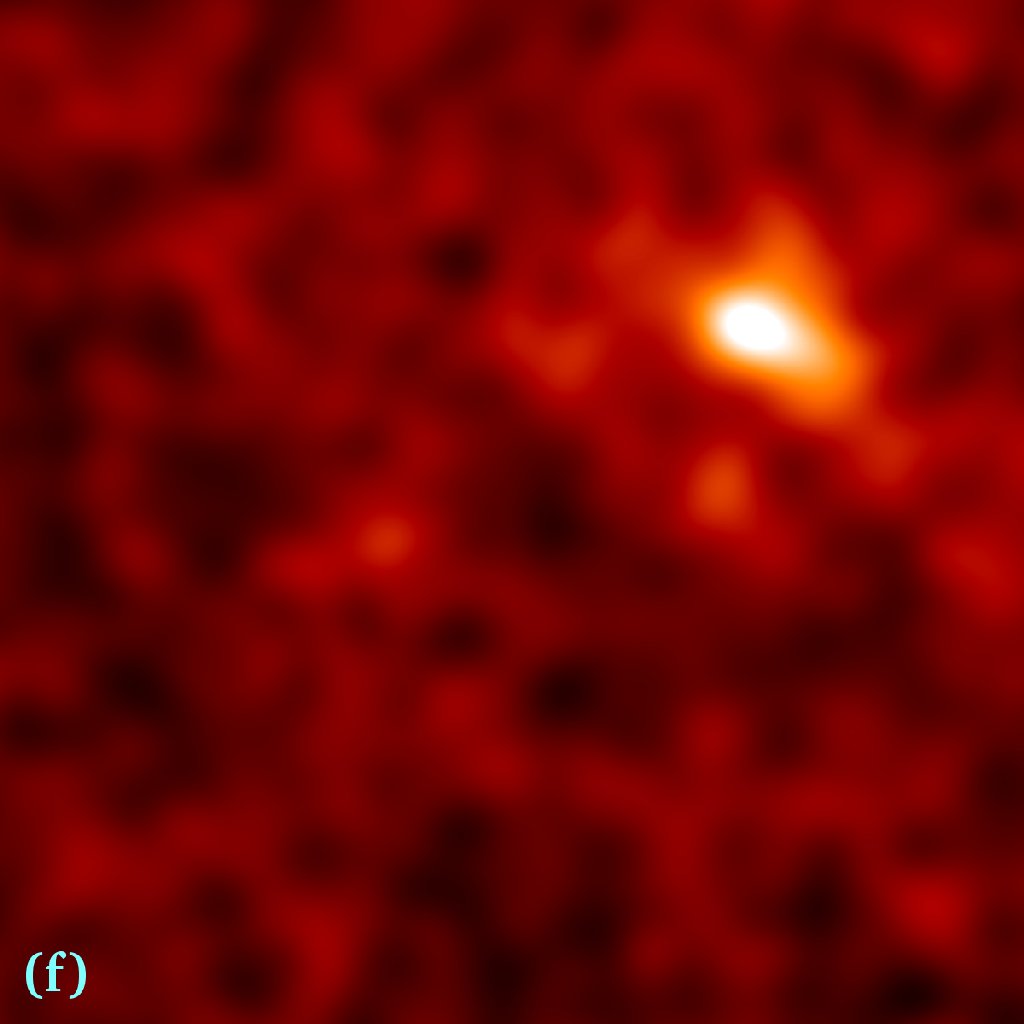}}
   \hspace{1em}
   \hspace{0.07\linewidth}
}
  \caption{
\label{fig:maps3}
Convergence maps smoothed at $\lambda=1\arcmin$ for the $20\arcmin\times20\arcmin$ field shown in Fig.~\ref{fig:maps1}:
(a) and (b) \21cm-based map \emph{without} and \emph{with} noise, resp.;
(c) and (d) space-based galaxy lensing map without and with noise;
(e) and (f) ground-based galaxy lensing map without and with noise.
All panels use the colour scale indicated by the bar at right.
}
\end{figure*}

\begin{figure*}
\centerline{\includegraphics[width=\linewidth]{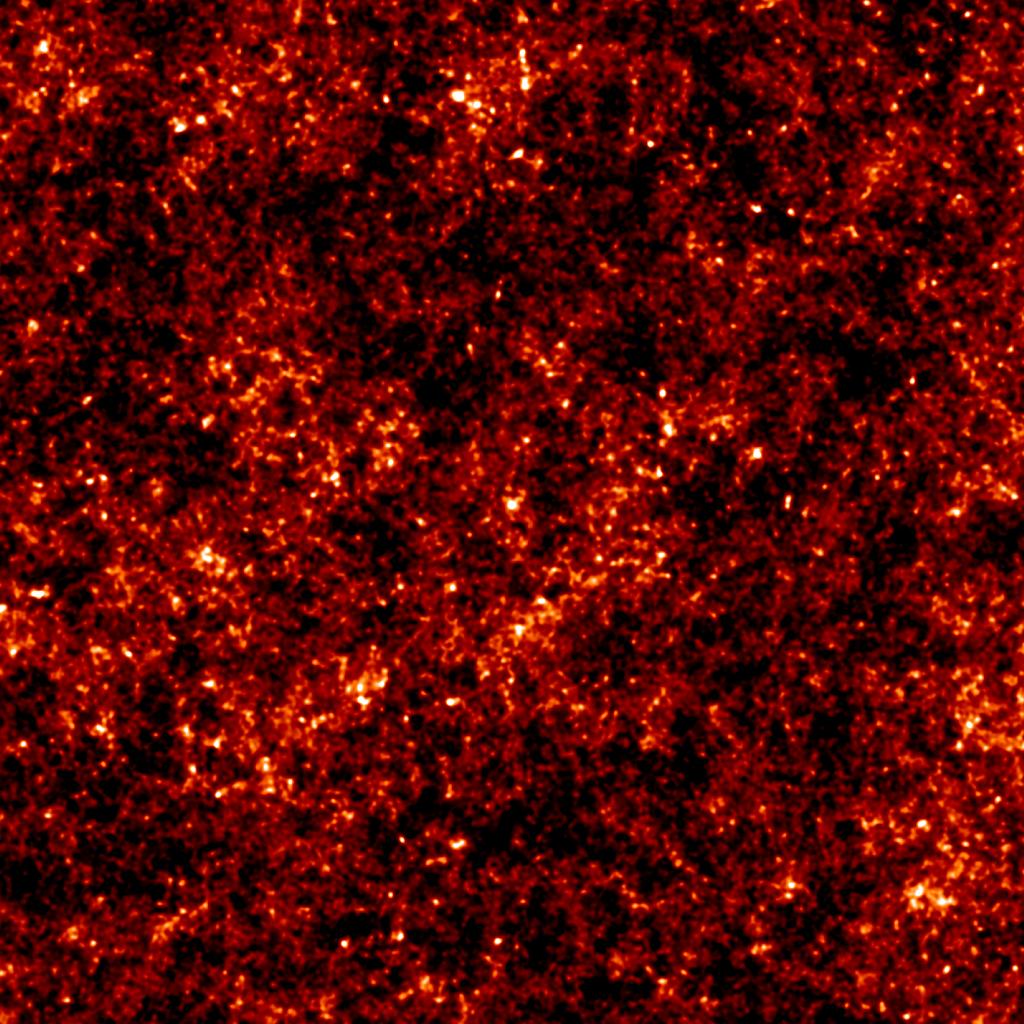}}
  \caption{
\label{fig:map4}
\21cm-based convergence maps for a $5\deg\times5\deg$ field with a  
$\lambda=1\,\arcmint$ beam and noise. The colour scale is the same as  
in Fig.~\ref{fig:maps3}. The field shown in Figs.~\ref{fig:maps1}, \ref{fig:maps2}, and \ref{fig:maps3} lies at the centre of this map.
}
\end{figure*}

Figures~\ref{fig:maps1}, \ref{fig:maps2}, and \ref{fig:maps3} show simulated convergence
maps. The $20\arcmin\times 20\arcmin$ field shown in these examples is  
only a small fraction ($1/900$) of the full $10\deg\times10\deg$ map that we
simulated.  The latter is too large to be displayed in sufficient
detail. This particular field was chosen because it has a prominent
mass concentration at upper right that is large enough to be detected
in all the cases we investigate.  This halo is at redshift $z=0.09$
and has a virial mass $M_{200}=6.9\times 10^{14} h^{-1}\,\msun$. Thus
it represents a galaxy cluster similar to the Coma cluster.  For this
reason, the field is not typical. From our Millennium Simulation data,
we find that there is only a 3\% probability for a random field this
size to contain a halo more massive than $5\times 10^{14}
h^{-1}\,\msun$.

The second largest mass concentration visible in the field (at lower
left) is at redshift $z=1.9$ and has a mass of $1.5 \times 10^{14}
h^{-1}\,\msun$. This is also a relatively unusual event. From our
data, we expect one halo with $M_{200}>10^{14} h^{-1}\,\msun$
and $z>1.5$ per square degree, corresponding to a 12\% probability for
a $20\arcmin\times 20\arcmin$ field.  There are three more halos with
masses above $10^{14} h^{-1}\,\msun$ visible. The most prominent of
these (left centre) is at redshift $z=0.46$ and has a mass of $1.2
\times 10^{14} h^{-1}\,\msun$. On average, we expect about two such
clusters in each $20\arcmin\times20\arcmin$ field.

The three convergence maps in Fig.~\ref{fig:maps1} are made without smoothing
or added noise in order to illustrate the dependence on the redshift
distribution of the sources. The map at top left gives the expected
convergence distribution (at $1\arcsec$ resolution) for sources at  
$z\sim 12$, representing the case of high-redshift \21cm lensing. The map at
top right uses the same colour table but a different source redshift
distribution, that appropriate for an optimistic space-based galaxy
survey.  The principal impression in comparing the two is that there
is much less structure in the `galaxy' map.  This reflects the
lowering of the overall amplitude caused by the smaller source
distances in the galaxy case. Averaged over the full $10\deg\times10\deg$ area,
the {\it rms} value $\sigma_\kappa$ of $\kappa$ is 0.11 in the
\21cm map, but only 0.03 in the galaxy map
(see Tab.~\ref{tab:kappa_stats} for a summary of the statistical properties
of our simulated maps). The map at lower left repeats the
galaxy map, but now the contrast is enhanced by a factor of 11/3 so
that the colour range matches that of the \21cm map.  Displayed in
this way, the two maps look similar. Nearby objects such as the most
massive cluster appear stronger in the galaxy map, whereas more
distant objects appear stronger in the \21cm map.  There are a few
large structures that appear in the \21cm map, but are absent from the
galaxy map. These are objects that lie beyond the redshifts assumed
for the galaxies. In particular, the large mass concentration at
$z=1.9$ is clearly visible in the \21cm map, but is virtually absent
in the galaxy map. For the reader's orientation, the lower right map
indicates masses and redshifts for all halos in the field with
$M_{200}> 10^{13} h^{-1}\,\msun$.

The two maps in Fig.~\ref{fig:maps2} illustrate expectations
for \21cm lensing reconstructions based on a (futuristic) radio telescope with a
Gaussian beam of width $\lambda=6\arcsec$ (corresponding to baselines $\sim100\,\km$).
The left image includes beam-smearing but excludes noise. A comparison with the  
$1\arcsec$ map at the top left of Fig.~\ref{fig:maps1} shows that very little detail is lost, and over
the full $10\deg\times10\deg$  area the {\it rms} value of $\kappa$ is
reduced by only 9\% to 0.098. The right image also includes
noise assuming the irreducible level expected for observations with
the optimal band width for a telescope beam of this size ($\sim 0.05\,\MHz$).
This has virtually no effect on the image, demonstrating that
such a (very large) telescope could produce high resolution mass maps
with very high fidelity. All structures with $M_{200}\geq 10^{13} h^{-1}\,\msun$ (these are indicated in Fig.~\ref{fig:maps1}d) can be clearly identified out to high redshift, and even many smaller halos down to masses $M_{200}\simgt 10^{11} h^{-1}\,\msun$ are visible.
The signal-to-noise at the scale of the beam
is very high even in low density regions, so substantial departures
from optimal conditions could be tolerated without significant
degradation of the resulting maps.

Figure~\ref{fig:maps3} shows maps smoothed with a Gaussian of width
$\lambda= 1\arcmin$. The colour scale is the same in all of them
and differs from those of Fig.~\ref{fig:maps1} an Fig.~\ref{fig:maps2}.
Now, the resolution is similar to that 
obtainable with the planned Square Kilometer Array \nocite{metcalf&white2007}  (Metcalf \& White2007).
The top two images
are for sources at $z\sim 12$ with no noise (left) and with noise at
the irreducible level expected for observations at the optimal
bandwidth for this beam-size (right).  Again the fidelity of the image
is high (although some differences can be seen in low $\kappa$ areas),
and many of the more massive halos indicated in Fig.~\ref{fig:maps1}d are detected
individually. The middle row of
maps are for our optimistic space-based galaxy lensing case, again
without and with noise, while the bottom row gives the corresponding
maps for our optimistic ground-based survey parameters.  As at higher
resolution, one is struck by how little structure is visible in these
maps compared to the \21cm case. The {\rm rms} value of $\kappa$ over
the full $10\deg\times10\deg$ field is smaller by factors of 3 and 4 in
the space- and ground-based galaxy cases in comparison with the \21cm
case (see Table 1).

The fidelity of the `observed' (i.e. noisy) maps is low in the
galaxy lensing cases.  A few of the structures seen in the noiseless
maps are still visible in their noisy counterparts, in particular the
largest object, but many of the low-amplitude peaks in these maps are
due to noise. In effect, only the large cluster at $z=0.09$ is
unambiguously detected for both surveys, while the two
$10^{14}h^{-1}\,\msun$ halos at $z=0.46$ also stand out above the
noise in the `space-based' map. The larger halo at $z=1.9$ remains unseen.
The fidelity of these maps could be
improved by increasing the smoothing length, but this would be at the
expense of losing all the individual halos.  In practice, an adaptive
smoothing method such as a maximum entropy scheme would probably be
used in order to remove low-significance features. This would leave
rather little structure visible in our $20\arcmin\times 20\arcmin$  
patch, only the
highest peak in the `ground-based' case. Only a few percent of
fields this size would contain an object massive enough to be detected
with high significance in a survey with these parameters.  This
limitation is quite evident in current ground-based lensing surveys
\nocite{2006A&A...452...51S,2000MNRAS.318..625B,2000astro.ph..3338K,
2000A&A...358...30V,2000Natur.405..143W}({Semboloni} {et~al.} 2006; {Bacon}, {Refregier} \&  {Ellis} 2000; {Kaiser}, {Wilson}, \&  {Luppino} 2000; {Van Waerbeke} {et~al.} 2000; {Wittman} {et~al.} 2000).

In order to give a better impression of the reconstruction
capabilities of a radio telescope such as SKA, Fig.~\ref{fig:map4} 
shows the map of Fig.~\ref{fig:maps3}b expanded to show a $5\deg\times 5\deg$ field.
(Note that this is still only 1/4 of the full field we simulated.)
Current plans for the SKA should enable this resolution to be reached using the dense
`core' array, but the noise level in the convergence map will depend on the way in which reionisation 
proceeds.  If the number density of ionised bubbles is large and they persist for a significant 
fraction of the redshift range expected for SKA ($7\simlt z \simlt 13$) then noise levels 
nearly this good can be obtained in 90~days of integration time.  A more pessimistic scenario 
is that reionisation happens very suddenly and nearly uniformly. Even if this is the case, 
and reionisation occurs near $z\sim 7$, moderate fidelity maps at $1.5\,\arcmint$ resolution should be possible and the same noise levels as in Fig.~\ref{fig:maps3} should be attainable but on 
$3\,\arcmint$ scales. In the latter case, SKA maps will be more noisy than Fig.~\ref{fig:maps3} after 90~days of 
integration, although still of much higher fidelity than galaxy-based maps.

The conclusion of this section is that galaxy lensing surveys do not
provide sufficient signal-to-noise to image any but the most massive
individual dark matter structures, but that a very large
radio telescope could, in principle, provide high-resolution,
high-fidelity images of the cosmic mass distribution in which the
halos of individual massive galaxies and galaxy groups are visible.

\subsection{Pixel distributions}\label{pixel_pdfs}

Another useful way to represent the information in our simulated maps
is to plot the probability density function for the convergence, ${\rm
pdf}(\kappa)$, in the different cases.  For this we can use the full
$10\times 10$ degree field, rather than the smaller subfields
discussed in section~\ref{sec:images}. Quantitative statistics for all
these distributions are given in Table 1.

\begin{figure}
\center
\includegraphics[width=8cm]{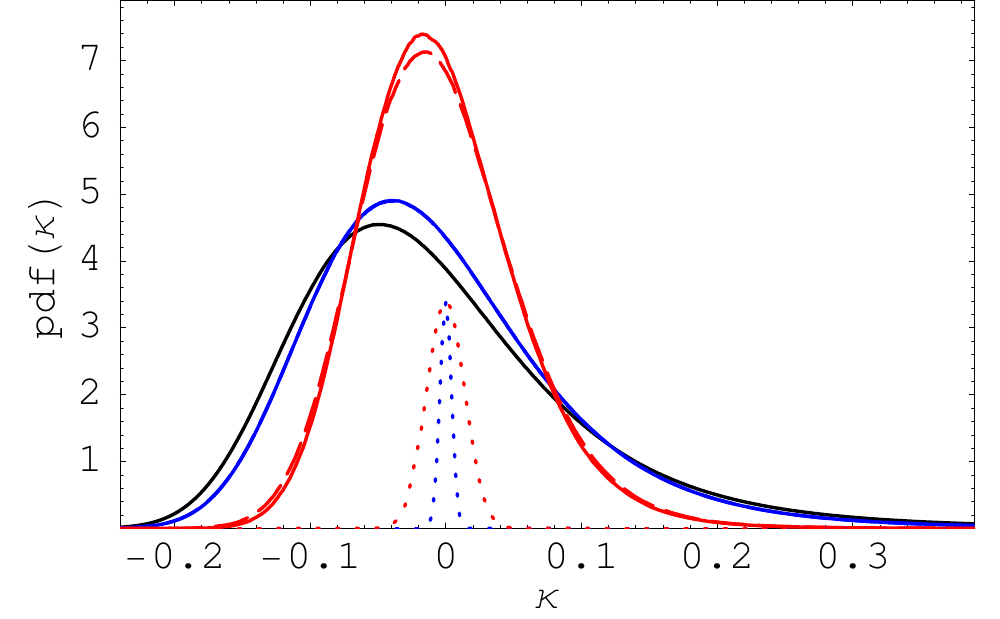}
\caption{
\label{fig:pdf_21cm}
The probability density function ${\rm pdf}(\kappa)$ of the
convergence for sources at $z=12$ (black line) compared to the
distribution smoothed with a Gaussian beam with $\lambda=6\arcsec$ (blue
solid curve), and $1\arcmin$ (red solid curve).  The dashed curves show  
the smoothed distributions with noise added at the irreducible level
expected for observations of the pregalactic HI with optimal total
bandwidth and frequency resolution.  The dotted lines illustrate the
noise distributions by themselves.  }

\end{figure}

Figure~\ref{fig:pdf_21cm} shows the results for sources at $z=12$, as
appropriate for pregalactic HI.  This confirms quantitatively our
previous conclusion that the irreducible noise has very little effect
on the maps.  Indeed, its effects are not even visible for a
$\lambda=6\arcsec$ beam, and they are still small for
$\lambda=1\arcmin$. This just reflects the fact that the pdfs for the
noise are much narrower than those for the signal, as illustrated in
Fig.~\ref{fig:pdf_21cm}.  Note that the narrower noise pdf is
associated with the higher resolution, higher amplitude signal pdf.

\begin{figure}
\includegraphics[width=8cm]{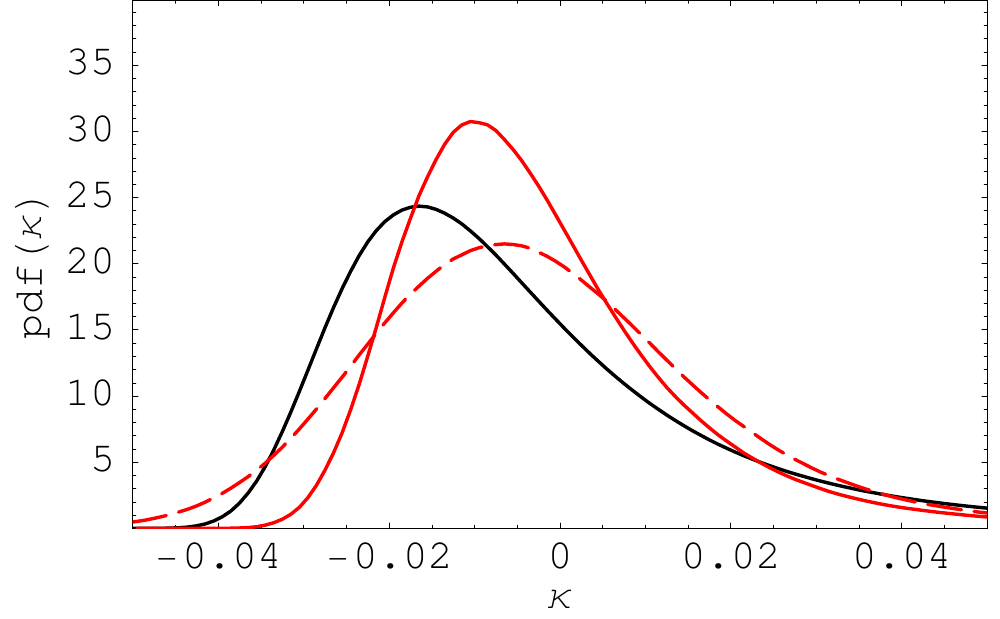}
\caption{
\label{fig:pdf_gals_space}
The probability density function ${\rm pdf}(\kappa)$ of the
convergence for an optimistic space-based galaxy lensing survey
(median redshift $z_\mathrm{med}=1.23$).  The unsmoothed, noiseless case
is shown in black.  The red curves are for Gaussian smoothing of width
$\lambda= 1\,\arcmint$.  The dashed and solid curves are with and
without noise respectively.  The assumed density of source galaxies is
$100\,\arcmint^{-2}$.  Note the difference in $\kappa$-scale compared
to Fig.~\ref{fig:pdf_21cm}.  }
\end{figure}

Figure~\ref{fig:pdf_gals_space} gives corresponding results for sources  
with the redshift distribution appropriate to a space-based galaxy lensing  
survey. Note that the scale has changed from Fig.~\ref{fig:pdf_21cm},  
reflecting the substantially lower amplitude of fluctuations in $\kappa$ in this case.
For reconstructions with a beam of width $\lambda=1\arcmin$, the noise  
expected in such a survey has a strong effect on ${\rm pdf}(\kappa)$. The low  
$\kappa$ tail of the observed distribution is practically all due to noise, and  
the shape of distribution is largely lost.  Estimating the skewness or  
kurtosis of the underlying distribution would clearly require a very good  
understanding of the properties of the noise.

\begin{figure}
\includegraphics[width=8cm]{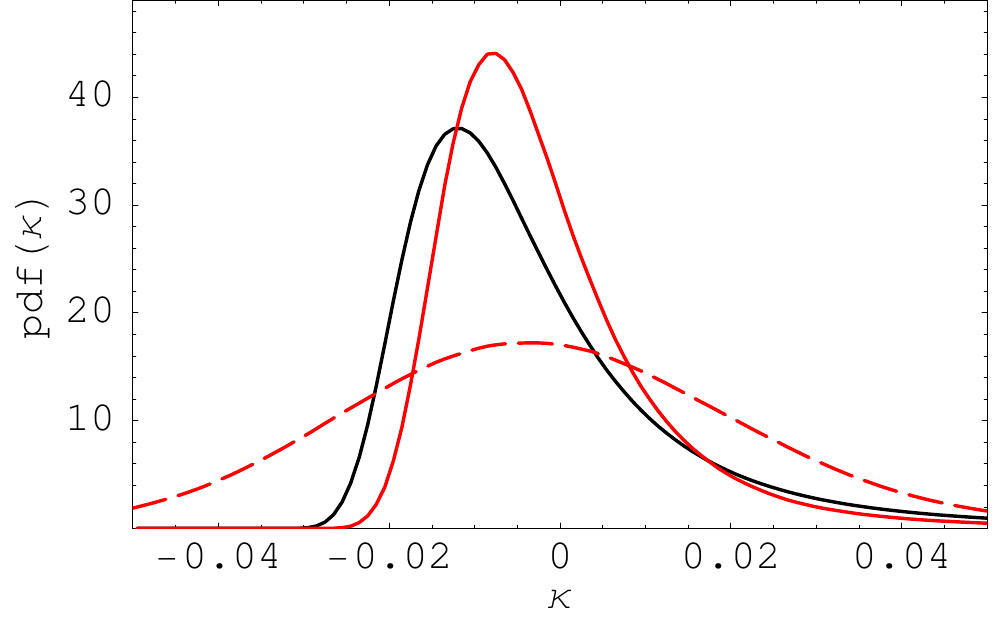}
  \caption{ \label{fig:pdf_gals_ground} The probability density
  function ${\rm pdf}(\kappa)$ of the convergence for an optimistic
  ground-based galaxy lensing survey (median redshift $z_\mathrm{med}=0.9$).  The unsmoothed, noiseless case is shown in black.  The
  red curves are for Gaussian smoothing of width $\lambda= 1\,\arcmint$.
  The dashed and solid curves are with and without noise
  respectively. The assumed source density is $35\,\arcmint^{-2}$.}
\end{figure}

The corresponding pdfs for the source
redshift distribution and noise appropriate to a ground-based galaxy  
survey are shown in Fig.~\ref{fig:pdf_gals_ground}.
Here the noise destroys almost all of the information in the original  
pdf.  With a large amount of data and with good knowledge of the systematics 
one can recover the variance accurately, but determination of higher moments
would be extremely challenging.

\begin{figure}
\includegraphics[width=8cm]{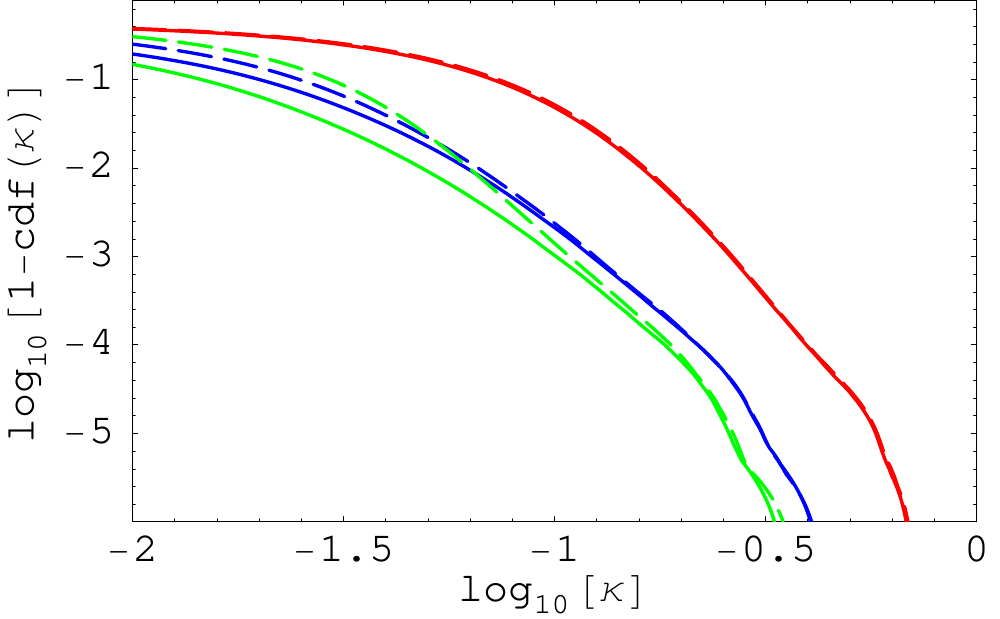}
  \caption{ \label{fig:cdf_comparison} The cumulative distribution
  function ${\rm cdf}(\kappa)$ of the convergence for HI at $z\sim 12$
  (red curves), for an optimistic space-based galaxy lensing survey
  (blue curves), and for an optimistic ground-based galaxy lensing
  survey (green curves). The solid curves represent the distribution
  with a Gaussian smoothing of width $\lambda=1\,\arcmint$ and with no
  noise. The dashed curves are the same but with noise included.  }
\end{figure}

Even when the noise is high compared to the dispersion in $\kappa$, it  
is still possible to measure the number density of very high mass objects.
Figure~\ref{fig:cdf_comparison} illustrates this point by plotting the  
high $\kappa$ tails of the cumulative distribution functions of $\kappa$.   
The noise has relatively little effect on these distributions for $\kappa  
\simgt 0.1$, even for the ground-based galaxy survey case. Such high values  
have a probability of around $\sim 10^{-3}$ corresponding to of order one  
object per square degree on the sky.  For our space-based survey parameters, the  
noise becomes unimportant for $\kappa\simgt 0.05$, corresponding to roughly  
100 objects per square degree.


\begin{table}
\center
  \caption{
\label{tab:kappa_stats}
The expected distribution of the convergence $\kappa$ for future
pregalactic HI, space- and ground-based galaxy surveys. For various
beam sizes $\lambda$ and noise levels $\sigma_\mathrm{N}$, we give the
standard deviation $\sigma_\kappa$, the skewness $S_3$, the $25\%$
quantile $\kappa_{25\%}$, and the $75\%$ quantile $\kappa_{75\%}$ of
the convergence distribution. (The mean of $\kappa$ is zero by
definition.)  }
\begin{tabular}{l l l l l l l}
\hline
\hline
Survey					& $\lambda$	& $\sigma_\mathrm{N}$	& $\sigma_\kappa$	& $S_3$	& $\kappa_{25\%}$	& $\kappa_{75\%}$\\
\hline
\21cm,					& 1\arcsec		& -					& 0.11			& 1.73	& -0.079			& 0.049	\\
$z_0=12$					& 6\arcsec		& -					& 0.098			& 1.35	& -0.071			& 0.046	\\
						& 6\arcsec		& 0.0042				& 0.098			& 1.34	& -0.071			& 0.046	\\
						& 1\arcmin	& -					& 0.058			& 0.52	& -0.044			& 0.031	\\
						& 1\arcmin	& 0.014				& 0.060			& 0.47	& -0.045			& 0.032	\\
\hline
space-based,				& 1\arcsec		& -					& 0.030			& 3.95	& -0.020			& 0.007	\\
$z_\mathrm{med}=1.23$,		& 1\arcmin	& -					& 0.018			& 1.99	& -0.014			& 0.006	\\
$n_\mathrm{g}=100\,\arcmint^{-2}$	& 1\arcmin	& 0.012				& 0.022			& 1.09	& -0.017			& 0.010	\\
\hline
ground-based,				& 1\arcsec		& -					& 0.021			& 4.61	& -0.014			& 0.005	\\
$z_\mathrm{med}=0.9$,		& 1\arcmin	& -					& 0.014			& 2.49	& -0.011			& 0.004	\\
$n_\mathrm{g}=35\,\arcmint^{-2}$	& 1\arcmin	& 0.02				& 0.024			& 0.31	& -0.018			& 0.014	\\
\hline
\end{tabular}
\end{table}

\section{conclusion}\label{sec:conclusion}

The noise in a mass map constructed using gravitational lensing of the
high-redshift HI distribution is expected to be much smaller than the
signal. In addition, the signal-to-noise {\it increases} with the
resolution of the map. It should thus be possible to make
high-resolution, high-fidelity images of the dark matter distribution
in which the dark halos of individual galaxies and galaxy groups are
visible.
For example, a very large future telescope with baselines $\sim 100\,\km$ may eventually allow us to detect halos with virial masses  $M_{200}\sim 10^{11}h^{-1}\,\msun$ out to redshift $z\sim10$ \nocite{metcalf&white2007} (Metcalf \& White 2007). Such detailed observations will provide a very direct and accurate test for structure-formation models. Even with an SKA-like telescope, halos with $M_{200}\simgt10^{13}h^{-1}\,\msun$ should be clearly detected out to high redshift.
This contrasts strongly with mass maps constructed using
gravitational lensing of distant galaxies, where high fidelity is only
achievable for angular smoothings so large that all but the nearest
and most massive individual objects are lost.

Our estimates of the irreducible noise are based on a convergence
estimator that is not necessarily optimal. It may therefore be
possible to achieve smaller `irreducible' noise levels than we
quote. In practice, however, it is likely that other sources of error
will dominate the overall budget, for example, the error introduced by
incomplete foreground subtraction \nocite{astro-ph/0608032}(see Furlanetto {et~al.} 2006, for a
review).  For most purposes, imaging the surface
density does not require reaching the irreducible noise limit; the
predicted signal is large enough to accommodate a noise level many
times the irreducible value.  In addition, the noise within a patch of
area $A$ goes down like $A^{-1/2}$ while the density fluctuations go
down roughly like $A^{-0.15}$, so even if the noise is too large to
map the surface density on the scale of a single beam, a high-fidelity
map with larger smoothing can still be constructed (as in the galaxy
lensing case).
The Square Kilometer Array in its currently proposed
configuration should be able to map the mass distribution on arcminute
scales with moderate to high fidelity if reionisation is not completed 
too early \nocite{metcalf&white2007}(Metcalf \& White 2007). The optimal bandwidth for observing lensing is $\sim 0.05\,\MHz$
while the signal-to-noise for mapping the pregalactic HI at the same angular scales is maximal at larger bandwidths $\sim 0.5\,\MHz$.
Lensing benefits from the stacking of many narrow redshift 
slices even if they are individually noise dominated while the temperature fluctuations 
themselves get diluted \nocite{metcalf&white2007}(Metcalf \& White 2007). To reach scales of a few arcseconds as 
discussed here will require a larger telescope with dense sampling. Given the narrower 
science goals, this may be achievable with simpler and cheaper antennas.

While high-resolution images of the cosmic mass distribution would be
a unique product of observations of \21cm lensing, they are not the
only reason to carry out such studies.  If enough of the sky can be
surveyed, cosmological parameters such as the density of dark energy
and its evolution with redshift can be measured with much higher
accuracy by a combination of \21cm lensing with galaxy lensing than
they can by galaxy lensing alone or indeed by any other method
proposed so far \nocite{metcalf&white2007}(Metcalf \& White 2007). The baseline configuration
of SKA may be powerful enough to achieve much of this improvement
if problems with foregrounds can be overcome.



\begin{thebibliography}{}

\bibitem[{Bacon}, {Refregier}, \&  {Ellis} 2000]{2000MNRAS.318..625B}
{Bacon}, D.~J., {Refregier}, A.~R., \& {Ellis}, R.~S. 2000, \mnras, 318, 625

\bibitem[Furlanetto, Oh, \&  Briggs 2006]{astro-ph/0608032}
Furlanetto, S., Oh, S.~P., \& Briggs, F. 2006, Phys.Rept., 433, 181

\bibitem[{Hilbert}, {White}, {Hartlap}, \&  {Schneider} 2007]{hilbert2007}
{Hilbert}, S., {White}, S.~D.~M., {Hartlap}, J., \& {Schneider}, P. 2007, \mnras, 382, 121

\bibitem[{Kaiser}, {Wilson}, \&  {Luppino} 2000]{2000astro.ph..3338K}
{Kaiser}, N., {Wilson}, G., \& {Luppino}, G.~A. 2000, ArXiv Astrophysics  e-prints, astro-ph/0003338

\bibitem[Massey {et~al.} 2007a]{Massey07b}
Massey, R. {et~al.} 2007a, \apjs, 172,239

\bibitem[Massey {et~al.} 2007b]{Massey07}
---. 2007b, \nat, 445, 286

\bibitem[Metcalf \& White 2007]{metcalf&white2007}
Metcalf, R. \& White, S. 2007, \mnras, 381, 447

\bibitem[{Semboloni}, {Mellier}, {van Waerbeke},  {Hoekstra}, {Tereno}, {Benabed}, {Gwyn}, {Fu}, {Hudson}, {Maoli}, \&  {Parker} 2006]{2006A&A...452...51S}
{Semboloni}, E., {et al.} 2006, \aap, 452, 51

\bibitem[{Smail}, {Ellis}, \&  {Fitchett} 1994]{1994MNRAS.270..245S}
{Smail}, I., {Ellis}, R.~S., \& {Fitchett}, M.~J. 1994, \mnras, 270, 245

\bibitem[{Springel}, {White}, {Jenkins}, {Frenk},  {Yoshida}, {Gao}, {Navarro}, {Thacker}, {Croton}, {Helly}, {Peacock}, {Cole},  {Thomas}, {Couchman}, {Evrard}, {Colberg}, \&  {Pearce} 2005]{SpringelEtAl05MSReview}
{Springel}, V., {et al.} 2005, \nat, 435, 629

\bibitem[{Vale} \& {White} 2003]{2003ApJ...592..699V}
{Vale}, C. \& {White}, M. 2003, \apj, 592, 699

\bibitem[{van Waerbeke} 2000]{2000MNRAS.313..524V}
{van Waerbeke}, L. 2000, \mnras, 313, 524

\bibitem[{Van Waerbeke}, {Mellier}, {Erben},  {Cuillandre}, {Bernardeau}, {Maoli}, {Bertin}, {Mc Cracken}, {Le F{\`e}vre},  {Fort}, {Dantel-Fort}, {Jain}, \& {Schneider} 2000]{2000A&A...358...30V}
{Van Waerbeke}, L., {Mellier}, Y., {Erben}, T., {Cuillandre}, J.~C., {et al.} 2000, \aap, 358, 30

\bibitem[{Wittman}, {Tyson}, {Kirkman},  {Dell'Antonio}, \& {Bernstein} 2000]{2000Natur.405..143W}
{Wittman}, D.~M., {Tyson}, J.~A., {Kirkman}, D., {Dell'Antonio}, I., \&  {Bernstein}, G. 2000, \nat, 405, 143

\bibitem[{Zahn} \& {Zaldarriaga} 2006]{ZandZ2006}
{Zahn}, O. \& {Zaldarriaga}, M. 2006, \apj, 653, 922

\bibitem[{Zwicky} 1933]{Zwicky33}
{Zwicky}, F. 1933, Helvetica Physica Acta, 6, 110

\bibitem[{Zaldarriaga}, {Furlanetto}, \& {Hernquist} 2004]{ZFH2004}
{Zaldarriaga}, M., {Furlanetto}, S. R., \& {Hernquist}, L. 2004, \apj, 608, 622

\end{thebibliography}
\end{document}